\documentclass[aps,prl,nobibnotes,twocolumn,superscriptaddress,bibliography]{revtex4-2}
\usepackage{amsfonts}
\usepackage{mathrsfs}
\usepackage{amsmath}
\usepackage{amssymb}
\usepackage{color}
\usepackage{natbib}
\usepackage{graphicx}
\usepackage{bm}
\usepackage{amssymb}
\usepackage{xspace}
\usepackage{epstopdf}
\usepackage{dcolumn}
\usepackage{longtable}
\usepackage{array}
\usepackage{makecell}
\usepackage{tabulary}
\usepackage{multirow}
\usepackage{appendix}
\usepackage{mathtools}
\usepackage{pifont}
\usepackage{longtable}
\newcolumntype{C}[1]{>{\centering\arraybackslash}p{#1}}
\newcolumntype{L}[1]{>{\flushleft\arraybackslash}p{#1}}

\usepackage{multirow}
\usepackage{epsfig}
\usepackage[normalem]{ulem}
\usepackage{amsmath}
\usepackage{braket}
\usepackage{bbold}
\usepackage{graphicx}

\makeatletter

\usepackage[colorlinks=true, letterpaper=true, pdfstartview=FitV, linkcolor=blue, citecolor=blue, urlcolor=blue]{hyperref}
\makeatother

\begin{document}

\title{Pseudo-Altermagnetism in Non-Hermitian Systems}

\author{Chang-Yin Ji}
\affiliation{State Key Laboratory of Chips and Systems for Advanced Light Field Display, Key Lab of advanced optoelectronic quantum architecture and measurement (Ministry of Education) and School of Physics, Beijing Institute of Technology, Beijing 100081, China}

\author{Runze Li}
\affiliation{State Key Laboratory of Chips and Systems for Advanced Light Field Display, Key Lab of advanced optoelectronic quantum architecture and measurement (Ministry of Education) and School of Physics, Beijing Institute of Technology, Beijing 100081, China}

\author{Zhi-Ming Yu}
\email{zhiming\_yu@bit.edu.cn}
\affiliation{State Key Laboratory of Chips and Systems for Advanced Light Field Display, Key Lab of advanced optoelectronic quantum architecture and measurement (Ministry of Education) and School of Physics, Beijing Institute of Technology, Beijing 100081, China}

\author{Yugui Yao}
\affiliation{State Key Laboratory of Chips and Systems for Advanced Light Field Display, Key Lab of advanced optoelectronic quantum architecture and measurement (Ministry of Education) and School of Physics, Beijing Institute of Technology, Beijing 100081, China}

\begin{abstract}
Altermagnetism has attracted growing attention, and its analogs have been actively explored across classical platforms. Here, we introduce pseudo-altermagnetism in non-Hermitian systems, where balanced gain and loss plays the role of the altermagnetic spin, and the opposite gain and loss sublattices are related by an antiunitary symmetry $O$ with $O\bm k\neq \bm k$ for generic momentum $\bm k$. Using a two-dimensional non-Hermitian system with $O=C_{4z}\mathcal{T}$ symmetry as a representative example, we find that pseudo-altermagnetism not only leads to a momentum-dependent gain and loss distribution analogous to the momentum-dependent spin splitting of altermagnets, but also exhibits intrinsic non-Hermitian properties. The bands of $C_{4z}\mathcal{T}$-invariant non-Hermitian systems are either isolated or paired by $C_{4z}\mathcal{T}$, and both cases are guaranteed to host movable but unremovable nodal structures, characterized by the zeros of two distinct quantities. An isolated band exhibits two movable but unremovable Hermitian spectral lines, whose generic points are not invariant under $C_{4z}\mathcal{T}$, in sharp contrast to the conventional Hermitian spectra enforced only at the fixed points of an antiunitary symmetry. The paired bands, in turn, feature a movable but unremovable nodal network formed by the union of real-part and imaginary-part degeneracy lines, whose intersections are generally exceptional points. We further demonstrate these findings in a non-Hermitian photonic crystal. Our results establish a symmetry-based bridge between non-Hermitian physics and altermagnetism, and provide novel mechanisms of Hermitian spectra beyond $\mathcal{PT}$ symmetry.
\end{abstract}

\maketitle

{\emph{\textcolor{blue}{Introduction.--}}}Altermagnets have recently emerged as a distinct class of magnetic order \cite{vsmejkal2022beyond,vsmejkal2022emerging,bai2024altermagnetism,zhang2024predictable,song2025altermagnets,fender2025altermagnetism,jungwirth2025altermagnetism}, combining the compensated magnetization of antiferromagnets with the momentum-dependent spin splitting of ferromagnets. Unlike conventional antiferromagnets, whose opposite-spin sublattices are related by translation or inversion, in altermagnets the opposite-spin sublattices are related by crystal rotations or mirror operations, making the spin splitting survive even without spin-orbit coupling. This unique combination gives rise to a variety of exotic phenomena in electronic systems 
\cite{zhou2024crystal,he2024quasi,gray2024time,ferrari2024altermagnetism,wang2025two,sun2025symmetry,gu2025ferroelectric,pan2026experimental,jungwirth2026symmetry,liu2026antiferroaxial,d2026altermagnetism,yuan2026quantum,cui2026altermagnetic,hu2026emergent,fu2026altermagnetism} and has inspired the search for altermagnetic analogs in other platforms.

Such analogs have been realized in photonic and acoustic crystals using  pseudospin degrees of freedom (DoF)  \cite{kim2026photonic,zhang2026photonic,cao2026mirror,chen2026altermagnetic,qiu2026orbital,xia2026time}. For example, altermagnetic photonic crystals have been constructed using chiral media \cite{kim2026photonic,zhang2026photonic,cao2026mirror}, as well as magneto-optical materials \cite{chen2026altermagnetic,qiu2026orbital}. Altermagnetic acoustic crystals have been realized using bilayer structures as  pseudospin-DoF \cite{xia2026time}. However, these approaches impose stringent requirements on material properties or involve complex sublattice coupling, making them difficult to extend to nonmagnetic, achiral and structurally simple systems.

\begin{figure}\includegraphics[width=0.4\textwidth]{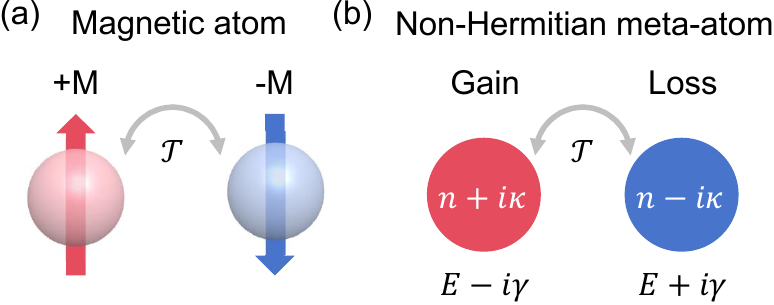}
\caption{Correspondence between altermagnetism and non-Hermitian systems. (a) A collinear altermagnet consists of two spin sublattices carrying opposite magnetic moments, $+M$ and $-M$. (b) A non-Hermitian lattice consists of two types of meta-atoms with relative permittivity (or refractive indices) $n+i\kappa$ and $n-i\kappa$, whose sign of $\kappa$ distinguishes gain from loss. 
The two binary degrees of freedom, $M$ and $\kappa$, are both odd under time reversal ${\cal T}$.}
\label{Fig-1}
\end{figure}

In contrast, non-Hermitian  systems  offer a natural alternative: the gain and loss form a binary DoF that changes sign under time reversal $\mathcal{T}$, just like spin. Non-Hermitian modifications of altermagnets have recently been investigated in electronic systems \cite{reja2024emergence,dash2025fingerprint,goswami2025investigation,ji2026non,zhao2026tailoring}. However, these works introduce gain and loss into real magnetic materials, whereas the complementary route---emulating the altermagnetic spin DoF with the binary gain and loss DoF---has remained unexplored.

Here, we introduce pseudo-altermagnetism in non-Hermitian systems, featuring at least one antiunitary symmetry $O$ whose spatial part is a nontrivial rotation or mirror with $O\bm k\neq \bm k$ for generic momentum  $\bm k$, such that the imaginary part $\gamma$ of the non-Hermitian spectrum satisfies $\gamma(O\bm k)=-\gamma(\bm k)$ and does not vanish identically. This requirement excludes antiunitary symmetry $\mathcal{PT}$, the symmetry relations of conventional antiferromagnets rather than altermagnets.

Focusing on two-dimensional (2D)  non-Hermitian systems with $O=C_{4z}\mathcal{T}$, we find that pseudo-altermagnetism not only yields a direction-dependent gain and loss distribution in momentum space, analogous to the direction-dependent spin splitting  in altermagnetic systems, but also exhibits intrinsic non-Hermitian properties. By symmetry analysis, we show that the bands in this system fall into two classes, isolated or paired by $C_{4z}\mathcal{T}$. Both cases exhibit symmetry-guaranteed movable but unremovable nodal structures. An isolated band hosts two movable but unremovable Hermitian spectral lines with $\gamma=0$, whose generic points are not invariant under $C_{4z}\mathcal{T}$, in contrast to the conventional Hermitian eigenvalues pinned only at antiunitary fixed points. The paired bands, in turn, support a movable but unremovable nodal network defined by $\operatorname{Im}\Delta_f=0$, where $\Delta_f=(\omega_1-\omega_2)^2$ is the two-band discriminant \cite{zhang2023symmetry}, corresponding to the union of the real-part and imaginary-part degeneracy lines, whose intersections are generally exceptional points (EPs). Here, $\omega_n$ is the complex eigenfrequency of $n$-th band.

We further demonstrate these findings in a non-Hermitian photonic crystal. Our results deepen the understanding of the interplay between the gain and loss DoF and crystal symmetries.

{\emph{\textcolor{blue}{Correspondence and distinction between gain-loss and spin.--}}}
In an altermagnet without spin-orbit coupling, the magnetic moment is a binary DoF with two opposite orientations, $+M$ and $-M$, which are distinguished only by their sign [see Fig.~\ref{Fig-1}(a)]. A non-Hermitian system carries an analogous binary DoF, gain and loss, distinguished by the sign of the imaginary part of relative permittivity or refractive index $\kappa$ [see Fig.~\ref{Fig-1}(b)]. Since both are odd under time reversal $\mathcal{T}$, gain and loss can be identified with pseudospin-up and pseudospin-down, respectively.

The correspondence is sharpened by the antiunitary symmetry $O$ relating the two sublattices. In an altermagnet, the spatial part of $O$ is a nontrivial rotation or mirror, so that $O\bm k\neq\bm k$ for generic $\bm k$. Acting on the bands, $O$ flips the spin and exchanges the two spin bands, $E_{\uparrow}(O\bm k)=E_{\downarrow}(\bm k)$. In the non-Hermitian counterpart, $O$ connects the two sites with opposite $\kappa$, and flips the sign of the  imaginary part of complex eigenfrequency $\omega=E+i\gamma$, leading to $E(O\bm k)=E(\bm k)$ but $\gamma(O\bm k)=-\gamma(\bm k)$.

However, the electron spin is a spinor with $\mathcal{T}^2=-1$. Thus, the spin-up and spin-down bands always come in pairs in altermagnets. In contrast, the gain-loss  is a scalar with $\mathcal{T}^2=1$, and the antiunitary symmetry $O$ alone cannot guarantee that the bands come in pairs. The bands of a spinless non-Hermitian system are therefore generally isolated, and the paired case arises only under an additional  symmetry condition. This is a common issue in classical analogues of altermagnetism \cite{kim2026photonic,chen2026altermagnetic,zhang2026photonic,cao2026mirror,qiu2026orbital,xia2026time}.

\begin{figure*}\includegraphics[width=0.7\textwidth]{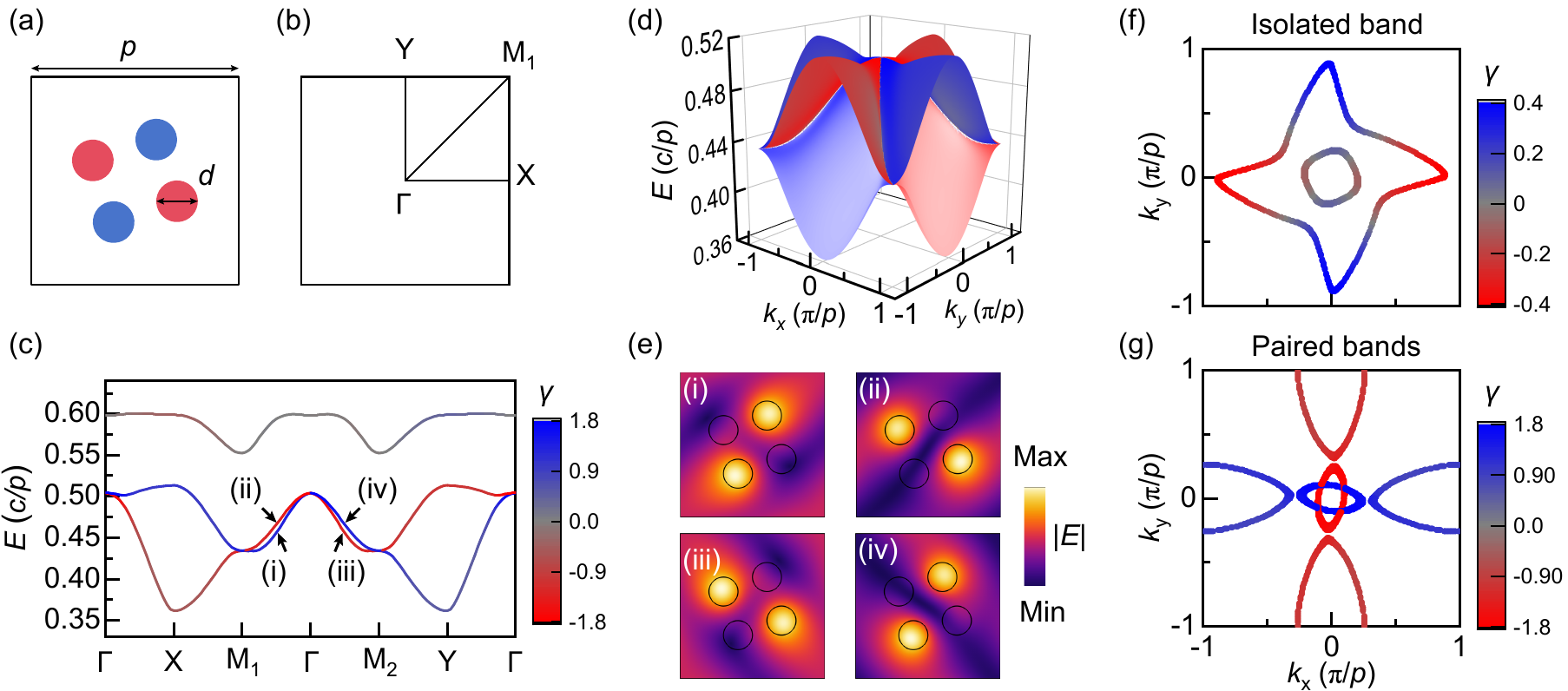}
\caption{Pseudo-altermagnetism in the non-Hermitian photonic crystal. (a) Schematic of the photonic crystal consisting of two pairs of dielectric cylinders with opposite gain (red) and loss (blue), with lattice constant $p$ and cylinder diameter $d$. (b) First BZ and the high-symmetry points. (c, d) Band structure along the high-symmetry path (c) and over the first BZ (d), with the color representing the imaginary part $\gamma$ of the eigenfrequency. Here, we set $d=0.2p$. The
gain cylinders are located at $(0.1p,0.2p)$ and $(-0.1p,-0.2p)$ with relative permittivities $12+i$, and loss cylinders at $(0.2p,-0.1p)$ and $(-0.2p,0.1p)$ with relative permittivities $12-i$.
(e) Electric-field intensity distributions $|E|$ for the four representative Bloch eigenmodes marked by (i)-(iv) in (c). (f, g) Pseudo-spin-resolved equifrequency contours of an isolated band at \(E=0.599c/p\) (f) and paired bands at \(E=0.502c/p\) (g). A time dependence of $e^{i\omega t}$ is assumed for all fields. The unit of $\gamma$ is $10^{-2}c/p$.}
\label{Fig-2}
\end{figure*}

An isolated non-Hermitian band is mapped to itself by $O$, whereas two bands related by $O$ are exchanged together with the sign flip of $\gamma$, leading to $\omega_1(O\bm k)=\omega_2^*(\bm k)$. Therefore, while altermagnets only realize the band-exchange case, because spin always appears as an up-down pair, the scalar gain and loss DoF admits both cases. Consequently, the non-Hermitian counterpart of the altermagnetic spin splitting appears only for the paired bands.

Moreover, for $O$-invariant momenta, the relations $\mathcal{T}^2=1$  and $\mathcal{T}^2=-1$ constrain the spectra in different ways. In altermagnets, the two spin bands are degenerate at the fixed points, where the spin splitting vanishes. For non-Hermitian systems, an isolated band is instead Hermitian with  $\gamma=0$, and the two paired bands have degenerate real parts, $E_1=E_2$, and opposite imaginary parts, $\gamma_1=-\gamma_2$, which generically do not vanish.

{\emph{\textcolor{blue}{Pseudo-altermagnetism with  $C_{4z}\mathcal{T}$ symmetry.--}}}
We now analyze a 2D non-Hermitian system with $C_{4z}\mathcal{T}$ symmetry. Because $C_{4z}\mathcal{T}$ is antiunitary, the spectrum at the $C_{4z}\mathcal{T}$-invariant momenta $\Gamma$ and M is classified by Wigner corepresentations rather than ordinary irreducible representations. Two types of corepresentations are possible: one-dimensional corepresentations, in which each band maps onto itself, satisfying $\omega(C_{4z}\mathcal{T}\bm k)=\omega^*(\bm k)$, and 2D corepresentations, in which two bands pair up and exchange, satisfying $\omega_1(C_{4z}\mathcal{T}\bm k)=\omega_2^*(\bm k)$. The two cases yield qualitatively different symmetry-guaranteed nodal structures, which we discuss in turn.

For an isolated band, $\gamma(C_{4z}\mathcal{T}\bm k)=-\gamma(\bm k)$ holds for every momentum $\bm k$. Thus, $\gamma$ is an odd continuous function on the Brillouin zone (BZ). The fixed points then give $\gamma(\Gamma)=\gamma(\mathrm{M})=0$, and hence Hermitian eigenfrequencies. In addition, the zero set of $\gamma$ must contain closed curves through $\Gamma$ and M. On a small $C_{4z}\mathcal{T}$-invariant circle around $\Gamma$, parametrized by the angle $\varphi$ with $C_{4z}\mathcal{T}$ acting as $\varphi\to\varphi+3\pi/2$, the relation $\gamma(\varphi+3\pi/2)=-\gamma(\varphi)$ forces at least four zeros on the circle. As the circle radius varies, these zeros trace curves through $\Gamma$, and likewise through M. These curves are the Hermitian spectral lines with  $\gamma=0$.

The Hermitian spectral lines can move with the system parameters but cannot be removed by any perturbation preserving $C_{4z}\mathcal{T}$. In particular, the generic points on the Hermitian spectral lines are not invariant under $C_{4z}\mathcal{T}$, in sharp contrast to the conventional Hermitian spectra enforce only at the fixed points of an antiunitary symmetry \cite{bender199,bender2024pt}.

Moreover, near $\Gamma$, the relation $\gamma(C_{4z}\mathcal{T}\bm k)=-\gamma(\bm k)$ forbids a linear term in $\gamma(\bm k)$. The leading order of $\gamma(\bm k)$ is therefore quadratic, $\gamma(\bm k)\approx A(k_x^2-k_y^2)+B k_xk_y$, the same $d$-wave form factor as in altermagnetic systems. Here, $A$ and $B$ are model-dependent real parameters. The Hermitian spectral lines near $\Gamma$ are thus the solutions of $A(k_x^2-k_y^2)+B k_xk_y=0$, and the same structure repeats around M.

For paired bands, the single-band argument fails, because $\gamma$ of each band is no longer an odd continuous function under $C_{4z}\mathcal{T}$. Instead, the symmetry maps the two bands onto each other, $\gamma_1(C_{4z}\mathcal{T}\bm k)=-\gamma_2(\bm k)$. At the fixed points, the paired bands have degenerate real parts and opposite imaginary parts, $E_1=E_2$ and $\gamma_1=-\gamma_2$, with the latter generically nonzero.
Since  $\omega_1(C_{4z}\mathcal{T}\bm k)=\omega_2^*(\bm k)$, we introduce a new $C_{4z}\mathcal{T}$-odd quantity that is well defined over the entire BZ,
\begin{equation}\label{Eq1}
\operatorname{Im}\Delta_f=2(E_1-E_2)(\gamma_1-\gamma_2),
\end{equation}
with  $\Delta_f=(\omega_1-\omega_2)^2$.
The zero set $\operatorname{Im}\Delta_f=0$ is the union of the real-part degeneracy lines $E_1=E_2$ and the imaginary-part degeneracy lines $\gamma_1=\gamma_2$. Since $E_1-E_2=0$  for $C_{4z}\mathcal{T}$-invariant momenta, the  real-part degeneracy lines pass through $\Gamma$ and M points. The union of the real-part and imaginary-part degeneracy lines is also movable but unremovable, by the same argument as for the isolated band. Near $\Gamma$, $E_1-E_2$ carries the same $d$-wave form factor. Since $\gamma_1-\gamma_2$ is even under $C_{4z}\mathcal{T}$, the imaginary-part degeneracy lines are generic curves that do not pass through the fixed points. The two families of lines intersect where $E_1=E_2$ and $\gamma_1=\gamma_2$ hold simultaneously, and these intersections are generally EPs at generic momenta.

{\emph{\textcolor{blue}{Photonic crystal demonstration.--}}}
To demonstrate pseudo-altermagnetism, we consider the non-Hermitian photonic crystal illustrated in Fig.~\ref{Fig-2}(a). Each unit cell contains four dielectric cylinders of diameter $d$ on a square lattice of period $p$. Two cylinders carry gain, with relative permittivity $12+i\kappa$, and two carry loss, with $12-i\kappa$. The total gain and loss are thus balanced, in analogy with the vanishing net magnetization of altermagnets. The gain cylinders sit at $(0.1p,0.2p)$ and $(-0.1p,-0.2p)$, and the loss cylinders at $(0.2p,-0.1p)$ and $(-0.2p,0.1p)$. This arrangement preserves $C_{4z}\mathcal{T}$, which maps each gain cylinder onto a loss cylinder, while breaking $\mathcal{PT}$.

The band structure along the high-symmetry path for $\kappa=1$ and $d=0.2p$ is shown in Fig.~\ref{Fig-2}(c), with the color encoding $\gamma$. We find that the bands indeed fall into two classes, isolated bands and paired bands. Here, the paired bands carry opposite $\gamma$, which changes sign between momentum sectors related by $C_{4z}\mathcal{T}$, forming an alternating gain-loss texture over the BZ [Fig.~\ref{Fig-2}(d)]. This is the non-Hermitian counterpart of the altermagnetic momentum-dependent spin splitting. The field distributions of the Bloch eigenmodes marked in Fig.~\ref{Fig-2}(c) are shown in Fig.~\ref{Fig-2}(e). The field distributions provide clear insight into the physical origin of the sign reversal of \(\gamma\). Moreover, the equifrequency contours of both the isolated and paired bands exhibit a $d$-wave feature [see Fig.~\ref{Fig-2}(f, g)], which is also a hallmark of altermagnetic systems.

\begin{figure}\includegraphics[width=0.46\textwidth]{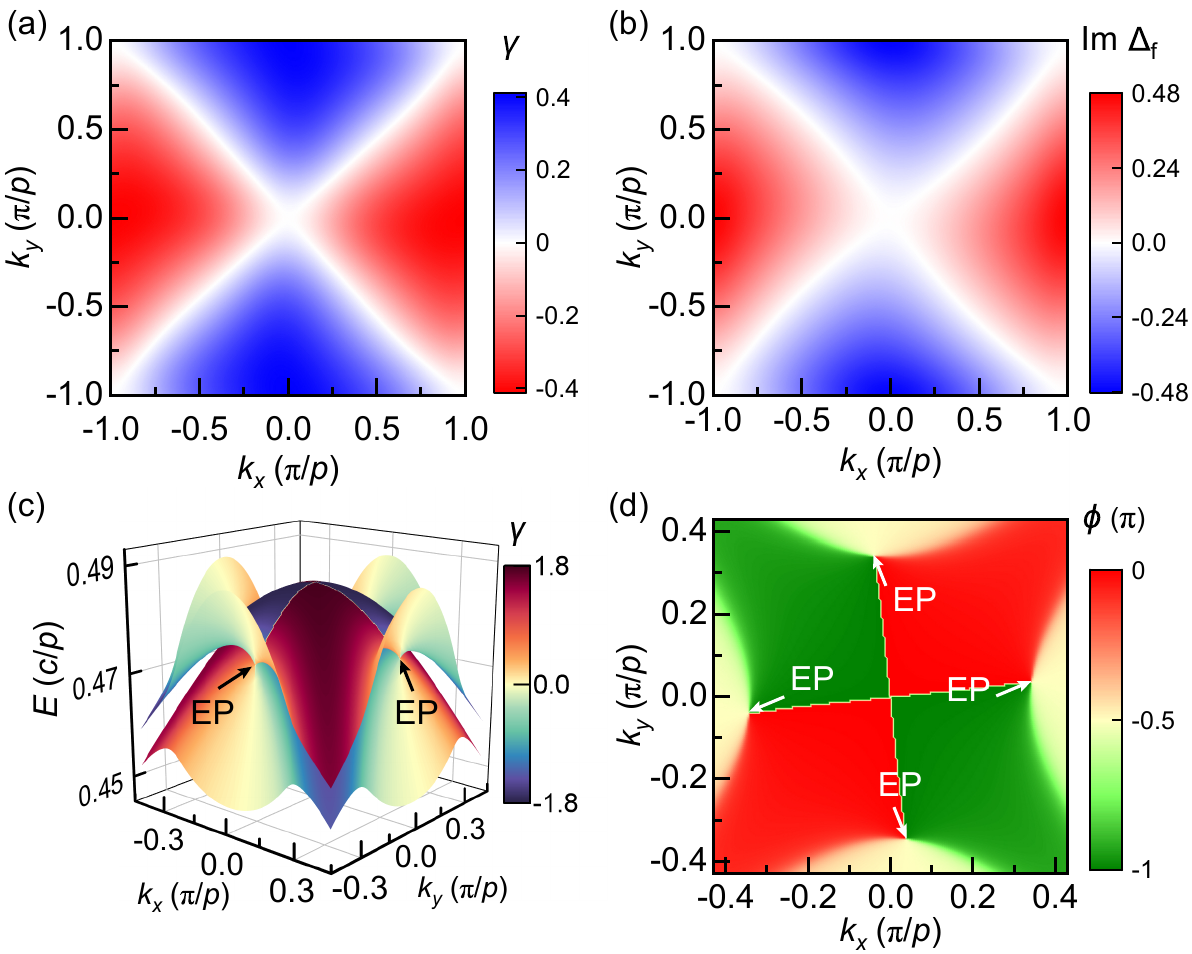}
\caption{$C_{4z}\mathcal{T}$-symmetry guaranteed nodal structures and exceptional points (EPs). (a) Distribution of $\gamma$ across the entire BZ for the isolated band shown in Fig.~\ref{Fig-2}(c). (b) Distribution of $\operatorname{Im}\Delta_f$ across the entire BZ for the paired bands shown in Fig.~\ref{Fig-2}(c). The unit of $\operatorname{Im}\Delta_f$ is $10^{-2}c^2/p^2$. (c) Band structures hosting EPs, with gain cylinders located at $(0.29p,0.2p)$ and $(-0.29p,-0.2p)$, and loss cylinders at $(-0.2p,0.29p)$ and $(0.2p,-0.29p)$. (d) Phase \(\phi=\arg[(\gamma_1-\gamma_2)+i(E_1-E_2)]\) for the band structure shown in Fig.~\ref{Fig-3}(c).}
\label{Fig-3}
\end{figure}

We now verify the two protected nodal structures predicted in the previous section. For an isolated band of this photonic crystal, Fig.~\ref{Fig-3}(a) shows $\gamma$ over the BZ. As shown, $\gamma$ varies continuously throughout the BZ, and its sign is opposite at momentum points related by $C_{4z}\mathcal{T}$, consistent with our analysis. As a consequence, two Hermitian spectral lines with $\gamma=0$ pass through $\Gamma$ and M. The two Hermitian spectral lines are curved, demonstrating that they can move within the BZ, yet they cannot be removed by any perturbation preserving $C_{4z}\mathcal{T}$.

For the paired bands, the protected quantity is $\operatorname{Im}\Delta_f$. The value of $\operatorname{Im}\Delta_f$ is plotted in Fig.~\ref{Fig-3}(b), where its zeros also form two curves passing through $\Gamma$ and M. Specifically, the two curves are the union of the real-part degeneracy lines $E_1=E_2$ and the imaginary-part degeneracy lines $\gamma_1=\gamma_2$. The real-part lines pass through $\Gamma$ and M, while the imaginary-part lines are generic curves that avoid the fixed points. 

More interestingly, by tuning model parameter, four EPs connected by $C_{4z}\mathcal{T}$  appear at the intersection of the  real-part and imaginary-part degeneracy lines, as shown in Fig.~\ref{Fig-3}(c). To identify these EPs, we define the phase $\phi=\arg\left[(\gamma_1-\gamma_2)+i(E_1-E_2)\right]$. At an EP, the real and imaginary parts of $\omega_1$ and $\omega_2$ are simultaneously degenerate, rendering $\phi$ ill-defined and giving rise to a phase singularity. As shown in Fig.~\ref{Fig-3}(d), the four EPs can therefore be identified as the singular points of the $\phi$ distribution. Each EP carries a topological charge of $q=\pm 1/2$, given by $q=\frac{1}{2\pi} \oint_C \nabla_{\mathbf{k}}\operatorname{\phi}(\mathbf{k})\cdot d\mathbf{k}$, where $C$ is a closed loop enclosing the EP. We further find that the $C_{4z}\mathcal{T}$ symmetry reverses the topological charges of the symmetry-related EPs [see Fig.~\ref{Fig-3}(d)].

\begin{figure}\includegraphics[width=0.45\textwidth]{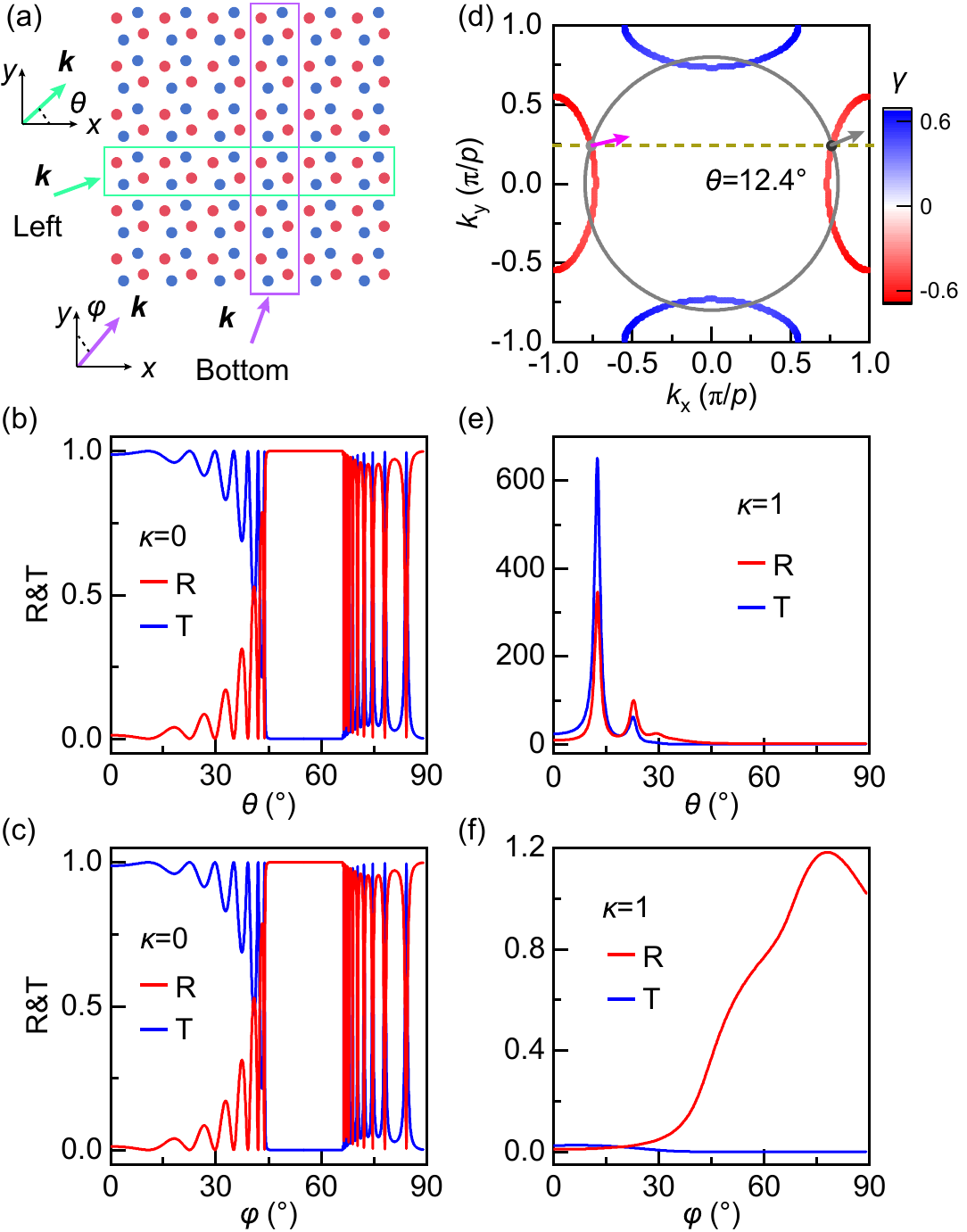}
\caption{Wave-matter interaction in pseudo-altermagnetism. (a) $\theta$ and $\varphi$ denote the incident angles measured from the $x$ and $y$ axes for incidence from the left and bottom sides, respectively. For left (bottom) incidence, periodic boundary conditions are imposed along the \(y\) (\(x\)) direction, with 27 unit cells along the \(x\) (\(y\)) direction. (b, c) Reflection $R$ and transmission $T$ spectra of the Hermitian photonic crystal ($\kappa=0$) for different incident angles and directions. (d) Pseudo-spin-resolved equifrequency contours at \(E=0.4c/p\). The gray circles indicate the light cones. The dashed line indicates the momentum corresponding to an incidence angle of \(\theta=12.4^\circ\) for light incident from the left. The gray and pink arrows indicate the energy-flow directions of the incident and outgoing waves, respectively, when the wave is incident from the left. (e, f) Reflection and transmission spectra of the non-Hermitian photonic crystal (\(\kappa=1\)) at \(E=0.4c/p\) for different incident angles and directions. The structural parameters are the same as those used in Fig.~\ref{Fig-2}(c).}
\label{Fig-4}
\end{figure}

We then examine how the alternating gain-loss texture manifests in wave scattering. A plane wave is incident from the left at an angle $\theta$ measured from the $x$ axis, or from the bottom at an angle $\varphi$ measured from the $y$ axis, as sketched in Fig.~\ref{Fig-4}(a). For each incidence, 27 unit cells are taken along the propagation direction, and periodic boundary conditions are imposed along the transverse direction. As a reference, Fig.~\ref{Fig-4}(b) and Fig.~\ref{Fig-4}(c) show the reflection and transmission spectra, denoted by $R$ and $T$, of the Hermitian crystal ($\kappa=0$) at $E=0.4c/p$ for the left and bottom incidences, respectively. The $R$ and $T$ spectra for the two incidences coincide, as required by the $C_{4z}$ symmetry, and satisfy $T+R=1$, with no amplification.

For the non-Hermitian crystal ($\kappa=1$), the $R$ and $T$ spectra in Fig.~\ref{Fig-4}(e) and Fig.~\ref{Fig-4}(f) become strongly angle-dependent, reflecting the momentum-dependent sign of $\gamma$ in Fig.~\ref{Fig-4}(d). The incident wave couples selectively to the Bloch mode that matches it in both frequency and transverse momentum. When the coupled mode is a gain mode with $\gamma<0$, energy is drawn from the gain medium, and both $R$ and $T$ may exceed unity. When it is a loss mode, the transmission $T$ is strongly suppressed.

Specifically, at $E=0.4c/p$, both $T$ and $R$ show pronounced peaks exceeding unity at $\theta=12.4^\circ$  for left incidence [see Fig.~\ref{Fig-4}(e)], where both the transmitted and reflected waves are gain modes, indicating strong amplification of both scattered channels. Notice that  the values of $R$ and $T$ are not determined solely by the sign of $\gamma$ of the scattered modes. They are also governed by the overlap between the incident wave and the coupled Bloch mode, which accounts for the variation of $T$ and $R$ at small incident angles even though the transmitted waves are all gain modes [see Fig.~\ref{Fig-4}(d)].

For bottom incidence, the transmission $T$ is suppressed at all incident angles [see Fig.~\ref{Fig-4}(f)]. At small angles, the transmitted waves are loss modes. At large angles, while the transmitted waves are gain modes, their group velocity along the $y$ direction is very small, preventing the amplified energy from reaching the far side. Similarly, the $R$ spectrum for bottom incidence is suppressed at angles where the reflected waves are loss modes, and recovers at large angles where the reflected waves are no longer loss modes. This angle- and side-selective amplification and attenuation, unavailable in Hermitian systems, constitutes the transport signature of the alternating gain-loss texture, offering a route to control wave amplification by incident direction and frequency.

{\emph{\textcolor{blue}{Discussion.--}}}In summary, we have introduced pseudo-altermagnetism in non-Hermitian systems, in which the binary gain and loss DoF behaves as a pseudospin. For a 2D $C_{4z}\mathcal{T}$-symmetric system, we find that both isolated and paired bands exhibit movable but unremovable nodal structures, which we have demonstrated, together with the angle-dependent wave amplification, in a non-Hermitian photonic crystal.

It is worth remarking that the correspondence between collinear compensated  magnetism and non-Hermitian physics is not complete even among the antiferromagnetic symmetries themselves. In electronic systems, both $\mathcal{PT}$ and $\mathcal{T}\tau$, where $\tau$ is the fractional lattice translation relating the two opposite-spin sublattices, are antiferromagnetic symmetries, and both enforce spin-degenerate bands. In non-Hermitian systems, however, the two symmetries give rise to qualitatively different spectra. An unbroken $\mathcal{PT}$ symmetry pins the entire spectrum to be real, whereas $\mathcal{T}\tau$ does not: for an isolated band, $\gamma(-\bm k)=-\gamma(\bm k)$ resulting from  $\mathcal{T}\tau$  merely enforces  Hermitian spectral lines through the $\mathcal{T}$-invariant momenta, and the gain and loss distribution on the equifrequency contours carries a linear ($p$-wave) form factor near $\Gamma$, in contrast to the $d$-wave texture of the pseudo-altermagnetism. 

The present framework is not restricted to photonics. Non-Hermitian phenomena are ubiquitous across electric circuits, acoustic, mechanical, and ultracold-atom systems \cite{xu2017weyl, ezawa2019non, stegmaier2021topological, bergholtz2021exceptional,yuan2022exceptional, cai2022exceptional, zhang2023observation, huang2024acoustic, bai2026recent}, all of which are spinless and nonmagnetic. Thus, pseudo-altermagnetism may be emulated in each of them with the same symmetry design. The predicted Hermitian spectral lines and $\operatorname{Im}\Delta_f=0$ networks can also be generalized to antiunitary symmetries beyond $C_{4z}\mathcal{T}$, and the direction-dependent wave amplification may find use in gain and loss engineering.

{\emph{\textcolor{blue}{Acknowledgments.--}}} This work is supported by the National Key R\&D Program of China (Grant 
No. 2025YFA1411201), and the National Natural Science Foundation of China (Grant Nos. W2511003, 12321004,  12204041, and 12474040).

\bibliography{ref}
\end{document}